\documentclass[12pt]{article}

\input{epsf}
\usepackage[dvips]{graphicx}
\usepackage{cite}
\def\1ad{\mbox{\normalsize $^1$}}
\def\2ad{\mbox{\normalsize $^2$}}
\def\3ad{\mbox{\normalsize $^3$}}
\def\4ad{\mbox{\normalsize $^4$}}
\def\5ad{\mbox{\normalsize $^5$}}
\def\6ad{\mbox{\normalsize $^6$}}
\def\7ad{\mbox{\normalsize $^7$}}
\def\8ad{\mbox{\normalsize $^8$}}

\def\npb#1#2#3{{ Nucl. Phys.} {\bf B#1} (#2) #3 }
\def\plb#1#2#3{{ Phys. Lett.} {\bf B#1} (#2) #3 }
\def\prd#1#2#3{{ Phys. Rev. } {\bf D#1} (#2) #3 }

\def\threedx{\mbox{$(3+\delta.x)$}} 
\def\dj{\hbox{d\kern-0.347em \vrule width 0.3em height 1.252ex depth
-1.21ex \kern 0.051em}}

\def\ket{\rangle}
\def\bra{\langle}

\def\bn{\overline n}
\def\bm{\overline m}
\def\bS{\overline S}
\def\bM{\overline M}
\def\bT{\overline T}

\def\bk{\overline k}
\def\bW{\overline W}
\def\bphi{\overline \phi}

\def\bbeta{\overline \beta}

\def\pt{\partial}

\newcommand{\be}{\begin{equation}}
\newcommand{\ee}{\end{equation}}
\newcommand{\ben}{\begin{equation*}}
\newcommand{\een}{\end{equation*}}
\newcommand{\ba}{\begin{eqnarray}}
\newcommand{\ea}{\end{eqnarray}}
\newcommand{\ban}{\begin{eqnarray*}}
\newcommand{\ean}{\end{eqnarray*}}
\newcommand{\brr}{\begin{array}}
\newcommand{\err}{\end{array}}
\newcommand{\bc}{\begin{center}}
\newcommand{\ec}{\end{center}}

\newcommand{\bea}{\begin{eqnarray}}
\newcommand{\eea}{\end{eqnarray}}
\newcommand{\bean}{\begin{eqnarray*}}
\newcommand{\eean}{\end{eqnarray*}}

\newcommand{\nn}{\nonumber }

\newcommand{\ie}{\mbox{\it i.e.~}}

\begin{document} 

\begin{titlepage}
\begin{flushright}
Saclay t00/118 \\
\end{flushright}

\vskip.5cm
\begin{center}
{\huge{\bf  Dilaton Stabilization in 
 \\ \vskip0.4cm Effective Type I String Models }}
\end{center}
\vskip1.5cm

\centerline{ S.A. Abel $^{a}$ and G. Servant $^{b}$  }
\vskip 15pt

\centerline{$^{a}$ LPT, Universit\'e Paris-Sud, Orsay,
91405, France}
\vskip 3pt
\centerline{$^{b}$ CEA-SACLAY, Service de Physique Th\'eorique,
F-91191 Gif-sur-Yvette Cedex, France.}

\vglue .5truecm

\begin{abstract}
\vskip 3pt

\noindent
We show that the dilaton and $T$-moduli can be stabilized by a single gaugino
condensation mechanism in the four-dimensional effective field 
theory derived from Type IIB orientifolds. 
A crucial role is played by the mixing of the blowing-up modes
$M_k$ with the $T$-moduli in the K\"ahler metric,
and by the presence of the $M_k$ in the gauge kinetic functions.
Supersymmetry breaking in these models is dominated by the 
auxiliary fields of the $T$ moduli, and phenomenologically 
interesting patterns can emerge.
\end{abstract}

\end{titlepage}

\section{Introduction}

Understanding how the dilaton gets a
phenomenologically consistent expectation value 
is one of the major 
problems of string-derived effective field theories. 
Perhaps the most promising approach
to dilaton stabilization is gaugino condensation
in some hidden gauge group, 
leading to the dynamical generation of a non-perturbative 
dilaton-dependent superpotential. 
In heterotic string theories, however, the 
simplest resulting scalar potentials do not 
stabilize the dilaton. Instead it 
runs away either to infinite values where the coupling
is weak, or to zero where the coupling becomes strong and 
perturbative control is lost.  

Attempts have been made to circumvent this problem by having 
a gauge group with several factors, and 
multiple gaugino condensation.
In this case, several exponential terms have to conspire to 
produce a minimum in the potential at finite dilaton values. 
These are the so-called `race-track' models. 
To be realistic, race-track models 
require that the gauge coupling at the 
string scale be compatible with estimates based on renormalization 
group evolution of the Standard Model gauge couplings. 
The vacuum expectation value of the dilaton is then constrained
to be $\langle Re(S) \rangle \sim 2 \sim g_{GUT}^{-2}$, 
which requires some degree of fine-tuning \cite{stringpheno,dine}.

In this paper we examine dilaton stabilization from gaugino condensation in 
effective theories of type I strings derived from type IIB orientifolds.
We find a picture that is radically different
from heterotic strings and in particular find that 
the dilaton can be stabilized with only one condensing gauge 
group. The novel feature of type I strings 
which allows us to do this is
the existence of twisted moduli, $M_k$, 
associated with fixed points. These not only modify the 
K\"ahler metric but also appear 
in the gauge kinetic functions, and consequently in the  
superpotential that is generated by gaugino condensation. 
As we shall see, it is the mixing of these new fields with 
the moduli in the K\"ahler metric which generically leads to a simple stabilization. 

After briefly presenting relevant aspects of Type I models, we
discuss in section 3 gaugino condensation and the dynamical 
superpotential that we will use in our study \footnote{we understand 
that gaugino condensation in type I theories
has been considered by Aldazabal, Font, Ibanez and Quevedo
(unpublished).}. Section 4 previews the
general features of the resulting scalar potentials in 
heterotic and type I models in order to 
explain why dilaton stabilization is considerably easier in the latter. 
In section 5 we give an explicit computation of the scalar potential 
in type I and describe a local minimum where the dilaton may 
be trapped. In section 6 we discuss the resulting soft breaking terms.

\section{Preliminaries; Structure of Type I Models}   

Type I string theories have interesting
phenomenological properties which have been 
investigated (using type IIB orientifolds) 
in refs.\cite{aiqu,imr,afiv,aiq}. 
For example, their brane structure allows the fundamental scale 
to be essentially a free parameter, and 
in addition the visible gauge couplings are no longer tied to the 
vacuum expectation value of the dilaton 
but can instead be determined by the twisted moduli fields. 
(Consequently the problem of stabilizing the 
dilaton and moduli is more democratic than 
in the heterotic case.)

In this paper we will be concerned with the effective 
scalar potential of type I models and the important aspects 
are therefore the gauge couplings 
and the K\"ahler potential which we now review.
The reader is referred to ref.\cite{orient} for details on 
the construction; ref.\cite{imr} for
a broad phenomenological outline, including 
the effect of choosing different fundamental scales; 
ref.\cite{pheno1,pheno1.5} for discussions of supersymmetry breaking 
and phenomenology with an intermediate fundamental scale;
ref.\cite{pheno2} for some other aspects of type I models.

Type I models constructed from type IIB orientifolds 
include different types of D-branes on which open strings can be attached
in various ways. Supersymmetric models either have just D9 branes or D9 
and D5 branes (by T-dualizing with respect to the three complex 
dimensions it is sometimes useful to exchange D9-branes with D3-branes and 
D5-branes with D7-branes). 
There are three classes of moduli fields that we need to consider:
the complex dilaton $S$, the untwisted moduli $T_i$ associated with 
the size and shape of the extra dimensions and the twisted 
moduli $M_k$ associated with the fixed points of the underlying orbifold.
In contrast with the Green--Schwarz mechanism of heterotic compactifications, 
the complex dilaton does not generically play any role in U(1) anomaly cancellation
of $D=4$, $N=1$ type IIB orientifolds. Instead, only the twisted 
moduli $M_k$ participate in the generalized Green--Schwarz 
mechanism~\cite{ano,iru}. 
Moreover, they induce a Fayet--Iliopoulos term which 
is determined by the VEVs of the $M_k$ fields and which can therefore be zero.
(In the heterotic case the FI term is given by the complex dilaton and 
is therefore constrained by the gauge couplings.)

In the gauge sector, gauge groups and charged chiral fields will 
depend on the type and location of D-branes present in the vacuum. 
One can generally consider the case with one set of
 9-branes and three sets of $5_i$-branes 
($i=1,2,3$). There are gauge groups $G_9,\, G_{5i}$ associated with each, 
and four types of charged matter fields;
$C_i^9$ ($i$ labels the three complex dimensions) comes from open strings 
starting and ending on the 9-branes; 
$C_i^{5_j}$ from open strings starting and ending on the same $5_i$-branes;
$C^{5_i5_j}$ from open strings starting and ending on different sets of 
$5_i$-branes; $C^{95_i}$ from open strings with one end on the 9-branes 
and the other end on the $5_i$-branes.

The gauge kinetic functions for a $Z_N$ orientifold model 
differ from the heterotic case.
Firstly there is no Kac--Moody coefficient multiplying the $S$-field 
dependence. In addition the blowing-up modes appear linearly, and 
for $G_{5i}$, the $S$-field is replaced by the $T_i$-fields.
For the D9-branes, the gauge kinetic function is \cite{iru,abd}
\be
\label{g9}
f_{9\, a} = S+\sum_k \sigma^k_a M_k \, ,
\ee
whereas for the D5-branes
\be
f_{5_i \, a}= T_i+\sum_k \sigma_{ia}^k M_k\, ,
\ee
where $\sigma_a^k$ are model dependent coefficients and
$k$ runs over the different twisted sectors.
The gauge coupling is given by $ {\rm Re} f_a=1/g_a^2$.

To describe the K\"ahler potential, we will
henceforth work with the overall modulus, taking $T_i = T$.
At one-loop level the K\"ahler potential for arbitrary numbers 
of $M_k$ fields has the form \cite{afiv},
\be
\label{Ka}
K = - \ln\, s - 3 \ln \tau + \hat{K}(m_k),  
\ee
where 
\be 
s = S + \bS \, ;\ \ \, \tau = T + \bT  - \sum_n|\phi_n|^2
\,;\ \ \, m_k = M_k + \bM_k - \delta_k \ln \tau\, ,
\ee

We have introduced generic fields $\phi$ to represent 
some linear combination of the $C^9_i$ or $C^{5\, j}_i$ fields which 
will later condense. (Formally,
our choice of putting the $\phi$ fields in a single $\tau$ 
corresponds to the linear combination $\phi_n=
\frac{1}{\sqrt{3}}(C^9_{1\, n}+C^9_{2\, n}+C^9_{3\,n })$. 
However it turns out that the minimization of the potential is 
independent of the particular linear combination to high order -- 
see later.)
They are singlets under the gauge group of the visible sectors but 
charged under the anomalous $U(1)_X$
and will appear in the dynamical superpotential. 
The first two terms are similar to the usual no-scale models \cite{ln}
where the $T$ and $\phi$-dependence appears in the combination $\tau$ only. 
Giving a VEV to $m_k$ takes us away from the orientifold point.
The $\delta_k \ln \tau$ term is a correction whose general form 
can be deduced from 
the one loop expression for the gauge coupling \cite{abd,lln} (and  
$\delta_k$ may be related to the
Green--Schwarz coefficients associated with $SL(2,R)$ anomaly cancellation
 \cite{lln}.)
There may also be dilaton dependent corrections to $m_k$ but 
their precise form is unclear (although various symmetry arguments
have been put forward for them~\cite{lln}) so we shall omit them,
assuming that they are negligible. 
(Note that the corrections in $m_k$ depend on the tree-level expression
for $K$. In contrast with previous work, we do not expand 
it by assuming small $\phi_n$, but instead retain the full 
no-scale structure, $\tau$.) 
For the moment we will also omit the various charged visible matter 
fields because they do not condense but will return to them 
later when we compute their soft masses. 

All that we currently know about the form of $\hat{K}$ is 
that it is an even function of 
$m_k$ thanks to the orbifold symmetry, and that 
the leading term in an expansion about the orientifold point,
$m_k=0$, is quadratic, $ \frac{1}{2} \sum_k m_k^2 $. 
Later we accommodate our ignorance by working with the parameter 
$x_k=\partial \hat{K}/\partial M_k $ where near the orientifold point 
$x_k\approx m_k$.

\section{Gaugino condensation in heterotic and type I}

In the heterotic string, a non-perturbative superpotential 
for the fields $S$ and $T$ can be generated by hidden sector
gaugino condensation with
gauge group $SU(N_c)$ and with extra `matter' in fundamental
representations. We shall consider only one 
flavour of quarks ${\cal Q}$ in the fundamental of 
$SU(N_c)$ and antiquarks $\tilde{{\cal Q}}$ in the antifundamental of 
$SU(N_c)$. Below the scale $\Lambda=e^{-f/2\beta}$, 
where $\beta$ is the
one-loop beta function coefficient of the hidden gauge group, 
the appropriate degree of freedom is the meson ${\cal Q} {\tilde{\cal Q}}$.
It is usual to treat the composite superfield, 
$ \phi_2=\sqrt{{\cal Q} \tilde{\cal Q}}$, as 
the relevant superfield appearing in the K\"ahler potential, 
and (for convenience) we will include it in $\tau $.
In addition to $\phi_2$ it will be necessary in both the heterotic 
and type I cases to include a field $\phi_1$ of 
charge $q_1$ in order to generate a perturbative 
mass term for $\phi_2$. 

The non-perturbative contribution to the superpotential 
can be fixed uniquely by considering global symmetries 
and reads~\cite{nonperturbative}
\be
W_{np} = 
\left( \frac{\Lambda ^{3 N_c-1}}{\phi_2^2 h(T) } 
\right)^{\frac{1}{N_c-1}}
\ee
where $h(T)$ is a product of Dedekind eta functions
resulting from a one-loop correction to $f$ (which
gives $W$ the required modular weight, -3)
and $\Lambda \sim e^{- k_N S/2 \beta}$.
(This is in the so-called `truncated' approximation; 
see ref.\cite{dent} for recent developments.)
Here $\beta = (3 N_c -1 ) / 16 \pi^2 $ and $k_N$ is
the Kac--Moody level of the hidden gauge group. 
Note that we have not yet `integrated out' 
any fields except the gaugino condensate.

In the heterotic string, the mixed
$U(1)_X\times \left[ SU(N_c) \right]^2  $ anomaly under the transformation 
\be
A_{\mu}^X(x) \rightarrow A_{\mu}^X(x) + \partial_{\mu}\alpha
\ee
is cancelled by the transformation 
\be 
\label{tmp1}
S\rightarrow S + \frac{i}{2} \delta_{GS}\alpha \, .
\ee
With one flavour, the anomaly is given by 
\be 
C_{N_c} = \frac{q_2}{2 \pi^2} = k_N \delta _{GS} \, ,
\ee
where $q_2=\frac{q+\tilde{q}}{2}$, is the $U(1)_X$ charge of $\phi_2$, 
and one can check that the total $W_{np}$ is invariant. 

The extension to type I models is straightforward.
Again we consider the
gauge group $SU(N_c)$ with one flavour of quarks ${\cal Q}$ in the 
fundamental of 
$SU(N_c)$ and antiquarks $\tilde{{\cal Q}}$ in the antifundamental of 
$SU(N_c)$, which together form a composite meson field, $\phi_2$.
Assuming that the $SU(N_c)$ resides on a D9-brane 
we now have $\Lambda=e^{-f_9/2\beta}$, where $f_9$ is given by
eq.(\ref{g9}). 
$W_{np}$ is fixed uniquely by global symmetries and reads
\be
\label{fix2aa}
W_{np} = \left( \frac{\Lambda ^{3 N_c-1}}{\phi_2^2}
\right)^{\frac{1}{N_c-1}}\, .
\ee
There is no $T$-dependence in this expression since
there is no $T$-dependence in the 
one-loop expression for the gauge kinetic function 
$f$ in the type I case. (If there exists a 
modular symmetry, the requisite modular weight of $W$ must therefore
come entirely from transformations of $M$.) 
The mixed anomaly under the $U(1)_X$ gauge transformation
is cancelled by a transformation of the $M_k$. Assuming only 
one $M_k = M$ we have
\be
M \rightarrow M + i \frac{\delta_{GS}}{2} \alpha
\ee
where $\delta_{GS}=\frac{C_{N_c}}{\sigma_{N_c}}=\frac{C_X}{\sigma_X}$. 
The $C_N$'s are the mixed anomaly
$U(1)_X\times[G_N]^2$ coefficients. Under
$U(1)_X$, $\Lambda$ has charge $ q_{\Lambda}=
\frac{\sigma_{N_c}}{2 \beta}\frac{\delta_{GS}}{2}=\frac{C_{N_c}}{4 \beta}$ 
and in our case 
$ C_{N_c}=\frac{q_2}{2 \pi^2}$ so that
$ q_{\Lambda}=\frac{2 q_2}{(3N_c-1)}$ and
again we see that $W_{np}$ is $U(1)_X$ invariant as required.

\section{Preview; heterotic versus Type I} 

Before presenting our results in detail, let us discuss in general 
terms why dilaton stabilization is difficult in the heterotic 
string, but can work in type I theories.
We first review the situation for heterotic strings
in the case where there is one condensing gauge group. We then 
preview the results (to be derived in later sections) 
for the scalar potential of effective type I theories, and highlight 
the new features that make a stabilization with a single gaugino 
condensate possible.
\\

\vspace{0.3cm}

\noindent\underline{\it The heterotic case}\\

\vspace{0.2cm}

\noindent 
Consider the effective theory for heterotic strings 
with $\delta_{GS}=0$ (so that $q_2=0$);
\be
K=-\ln\,s -3\ln \,\tau \, ,
\ee
where $\tau$ is as defined above, and includes hidden sector 
fields, $\phi_n$. The $F$-part of the supergravity scalar potential 
is given by
\ba
V_F &=& e^G (-3 + G_{\alpha} K^{\alpha \bbeta} G_{\bbeta}) \nn\\
&=& \frac{1}{s \tau^3} \left( |s W_S - W|^2 
+ \frac{\tau }{3} | W_n + \overline{\phi}_n W_T | ^2 
+ \frac{1}{3} | \tau W_T - 3 W |^2 - 3 |W|^2 \right)\, ,
\label{het0}
\ea 
where
\be
G=K+\ln|W|^2\, ,
\ee
and subscripts indicate differentiation. 
If $W$ does not depend on $S$ or $T$ then 
\be
V_F\sim \frac{|W|^2}{s\tau^3} \ \ \, 
\ee
and obviously neither $S$ nor $T$ are stabilized.

In order to attempt a stabilization we invoke a non-perturbative 
superpotential as described in the previous section. We also 
add an additional field, $\phi_1$ (which for this example we
take to have zero charge under $U(1)_X$),
which generates a mass term for the 
mesons. The effective superpotential contains a perturbative piece,
so that we can write
\be 
W = W_p + W_{np},
\ee
where $W_p$ includes a mass term for $\phi_2$; for example 
\be 
W_p = a \phi_1 \phi_2^2 + b \phi_1^3 \, .
\ee
(Note that more general functions of these 
invariants are possible but we restrict ourselves to 
the linear case here.) We also assume that the fields are uncharged
under all other symmetries so that we can ignore 
the $D$-terms for this example.

Now let us look at the minimization of the scalar potential in 
eq.(\ref{het0}). 
The usual assumption to make is that at the minimum the VEVs of {\em
all} the $\phi_n $ are much smaller than any of the moduli. 
The potential is therefore dominated by the $|W_n|^2$ terms and 
setting $W_n=0$ determines $\phi_1$ and $\phi_2$ in terms of $\Lambda$. 
For any reasonable value of $s$, the third term (involving 
$W_T$) fixes the VEV of the $T$ modulus to a value
close to $T=1.2$, upto modular transformations. 
After these minimizations the effective superpotential is
\be 
W_{eff} \sim \frac{e^{\frac{-3 k_N S}{2 {\beta}}}}{\eta(T)^6} \, .
\ee
This is the effective potential after `integrating out' the 
mesons and the $\phi_1$ field, and is often the starting point for 
studies of dilaton stabilization.

The remaining dilaton dependence in
the scalar potential can then be written,
\be
\label{het}
V_F \sim \frac{e^{-2 \Delta s}}{s}(g+(1+\Delta s)^2)
\ee
where $\Delta =-W_S/W= \frac{3 k_N}{2 \beta}$ is a {\it positive}
constant, and $g$ is independent of $s$.
The point to appreciate here (because it will contrast with 
the type I case) is that $\Delta$ is fixed as soon as 
we eliminate the $\phi_n$ fields using the $W_n=0$ constraint.

\begin{center}
\begin{figure}[htb]
\vspace*{-0.6in}
\epsfxsize=7.85in
\epsffile{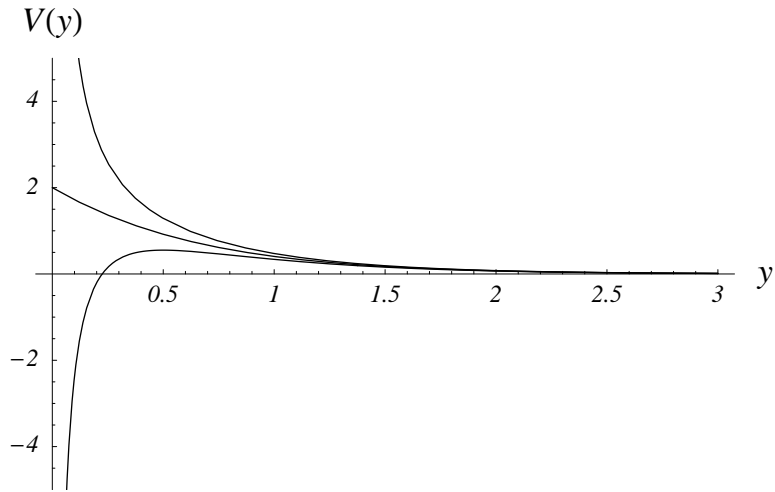}
\vspace{-6.6in}
\caption{$F$-part of simple heterotic-type potentials 
(see expression(\ref{het})). $y$ is proportional 
to the real part of the dilaton.}
\label{v1}
\end{figure}
\end{center}

Defining $y=\Delta s$, the minimization condition is
\be
(1+y)(1+g)+y^2+y^3=0\, .
\ee
This leads to the following 
situations (see figure \ref{v1}): If $g \geq -1$
there are no positive solutions. If $g < -1$, 
there is one positive solution to this equation
which is a maximum, with 
the potential running to zero at infinite $s$ and $-\infty $ at $s=0$. 
In all cases there is no minimum at positive and finite $s$.

As mentioned in the introduction, 
`race-track' models get around this problem by
considered several asymptotically free gauge groups
(see for example reference ref.\cite{stringpheno}).
$W$ can then be a sum of exponentials which can 
conspire to give rise to a
local minimum with non vanishing gauge coupling.
There are two other approaches that have been taken in the 
context of heterotic string theory. The first also requires several
group factors, but assumes that one of them is not asymptotically free;
\ie it has negative $\beta$. This contribution to the superpotential 
removes the minimum at $s \rightarrow \infty $, and the stabilization 
occurs rather more naturally~\cite{quevedoetal}. 
The second approach \cite{kahlstab} is to assume that the
K\"ahler potential receives non-perturbative corrections of the 
form $e^{-1/g}\sim e^{-\sqrt{{\rm Re} S}}$ 
as first conjectured by Shenker~\cite{sh}
\footnote{Note that this``K\"ahler stabilization" would not be 
possible in type I models 
where the divergence of perturbation theory in open strings
indicates only $e^{-1/g^2}$ terms~\cite{que}.}.
We should mention at this point a completely different way of 
generating a non perturbative superpotential for $S$ which was 
noted in ref.\cite{dg}. 
By compactification of M-theory using a Scherk--Schwarz
mechanism, the authors obtained a superpotential linear in $S$ whose 
minimization in the absence of matter gives $S=1$. \\

\vspace{0.3cm}

\noindent\underline{\it The type I case}\\

\vspace{0.3cm}

\noindent We now contrast the above 
with the general situation that 
we will find in type I theories in the following sections. 
We will consider the scalar potential with a  
K\"ahler potential given by eq.(\ref{Ka}) and with only one $m_k$ which we 
call $m$. We will assume a single gaugino condensate
in a hidden sector living on a D9-brane
which generates a $W_{np}$ as described 
in the previous section. The superpotential may be written
\be
\label{sup}
W=W_p(\phi_1,\phi_2)+W_{np}(\phi_2,S,M) \, .
\ee
There are two important differences with respect to the heterotic case. 
The first is that there are no factors of $\eta(T)$ appearing in the 
superpotential, and therefore $W$ does not necessarily depend on $T$
if the condensing group lives on the D9-brane. The second difference 
is that $W_{np}$ depends on the gauge kinetic function 
$f_9 = S+\sigma M $ and therefore includes both $S$ and $M$. 

We will find $F$-term contributions of the form 
\be
V_F = e^G B 
\ee
where 
\be 
B=  g(m) + 
|1 + s \Delta|^2 + \frac{\tau }{3} \left| \frac{W_n}{W} \right| ^2 
+ | B_0(m) - \sigma \Delta |^2 \, ,
\label{typeI}
\ee
$g$ and $B_0$ are functions of $m$ only, and where we have 
assumed $W$ is independent of $T$.

The most important aspect of this expression is that $\Delta $ appears 
{\em twice} because  $W_{np}$ is a function of $f_9$ and hence
\be 
\frac{W_M}{W}= \sigma \frac{W_S}{W} = - \sigma \Delta\, .
\ee
Now let us demonstrate the existence of 
a minimum when both $g(m)$ and $B_0/\sigma$ are small and negative.
Provided $g$ is small enough (which we check is always the case) or zero, 
we can neglect the $e^G$ prefactor and discuss the minimization of $B$.
For a given $m$, the minimum of $B$ is close to where the 
squared terms all vanish. However, gauge invariance of $W$ tells 
us that 
\be
q_n \phi_n \frac{W_n}{W} 
- \frac{\delta_{{GS}_k}}{2} \sigma \Delta = 0\, ,
\ee
and so $W_n=0$ cannot be satisfied at the same time as 
$B_0(m) - \sigma \Delta =0$ if $B_0\neq 0$. 

We therefore 
have a different option to the heterotic case. To simplify the 
discussion in this section, suppose that  
\be 
\label{cond1}
\sqrt{\frac{\tau}{3}} 
\left| \frac{1}{W}\frac{\partial W_{np}}{\partial \phi_2} \right|
\ll \left| \sigma\Delta \right| 
\, , \left| s\Delta \right| \, .
\ee
In this case it is clear that the minimum will be near 
\ba 
\label{fix}
\frac{\partial W_p}{\partial \phi_n} &=&0 \nn\\
B_0(m) - \sigma \Delta &= & 0 \, .
\ea
Eq.(\ref{cond1}) is easily satisfied if the additional
$D_X$ terms in the potential generate a large VEV for $\phi_2/\sqrt{\tau}$ 
whilst the VEVs of the $\phi_{n\neq 2}$ remain small.
If the $\phi_n$ fields are charged under $U(1)_X$, setting $D_X=0$ gives 
\be
\frac{q_2 |\phi_2|^2}{\tau} \approx 
\frac{ |\delta_{GS} K' |}{6} \approx
\frac{ |\delta_{GS} m |}{6} \, . 
\ee 
The nett result of eq.(\ref{cond1}) is a lower bound 
on the Fayet-Illiopoulos term, so that $m$ cannot be zero.

We stress that the main difference with the heterotic case 
is that here $\Delta $ is fixed by eq.(\ref{fix}) rather than 
being fixed (implicitly) by $W_2=0$. Indeed, we find
\be 
\Delta = \frac{B_0}{\sigma} \,\, ; \, s = - \frac{\sigma}{B_0} \, ,
\ee  
provided that $B_0/\sigma $ is negative, thereby fixing 
both the $\phi_2$ condensate and the dilaton. 
Note that the minimum at $s\rightarrow \infty $ still exists, but 
if for example $g(m)$ is small and negative the minimum at finite $s$ 
is lower. Indeed we can eliminate $\Delta $ to find 
\be 
B = g + \frac{(\sigma+B_0 s )^2 }{(\sigma^2 + s^2 )} \, .
\ee
A typical potential (with $g(m) = 0$) is plotted
in fig.(\ref{v3}) (including the $1/s$ prefactor from $e^K$). 
(When $g<0$ the minimum is at negative values of $V$.)

Since $\phi_2/\sqrt{\tau}$ is already fixed by the $D_X$ term
the remaining task is of course to show that $g(m) $ can actually have 
a minimum at small negative values, and that at this point 
$B_0/\sigma$ is indeed negative. One of the main results of the
explicit discussion in the sections which 
follow is the condition on $\hat{K} $ required to form 
a minimum for $m$ close to $m=0$ with small and 
negative cosmological constant.
We also discuss a different possibility 
which is reminiscent of the `no-scale' models. We can 
tune the cosmological constant to be exactly zero 
by fine-tuning $\sigma$. In this case the value of 
$m$ is undetermined and instead 
negative $\delta_{GS} m$ values parameterize a flat direction with stabilized
dilaton VEV.

The detailed discussion in the following sections
will demonstrate that the behaviour we have outlined above is very general.
Indeed we find that the dilaton is stabilized 
even when the inequality in eq.(\ref{cond1}) is not satisfied,
and hence the only requirement is that $W$ is a function of $f_9$ 
and that $B_0/\sigma<0$. Finally we consider the 
pattern of supersymmetry breaking which emerges. 

\section{Minimization of the SUGRA Scalar Potential}

We now compute the scalar potential using the K\"ahler (\ref{Ka}) and the
superpotential (\ref{sup}). 
We get the following for the first derivative of the K\"ahler potential:
\be 
K_\alpha = \left( 
-\frac{1}{s}
\, , \, 
- \frac{ (3+\delta . x) }{\tau}
\, , \, 
x_k 
\, , \, 
\bphi_n  \frac{ \threedx }{\tau} \right) \, . 
\ee
Here we have introduced $x_k = \pt \hat{K} /\pt M_k $, and 
have defined a dot product, $\delta . x \equiv \sum \delta_k x_k $.
In the small $m_k$ limit we have $x_k \approx m_k$.
The fact that the $K_n$ terms are $-\bphi_n\times K_T$ 
(which is really a result of the `no-scale' structure) 
will make the scalar potential simplify dramatically. 
Differentiating again we get 
\be 
K_{\alpha\bbeta} = \left( 
\begin{array}{cc}
\frac{1}{s^2} & 0 \nn\\
0 & K_{a\overline b} 
\end{array} \right) .
\ee 
To express $K_{a\overline b}$ we define 
\ba 
J_{kk'} &=& \frac{\pt^2 \hat{K} }{\pt M_k \pt \bM_{k'} }\nn\\
A_{kk'} &=& (J^{-1})_{kk'} \threedx + \delta_k\delta_{k'} \nn\\ 
C       &=& \threedx + \delta.J.\delta \nn\\
\delta.J.\delta &=&\delta_k J_{kk'} \delta_{k'}
\ea
and then have 
\be 
K_{a\overline b} = \left( 
\begin{array}{ccc}
\frac{C}{\tau^2} & -\frac{(\delta.J)_k}{\tau} & -\phi_m \frac{C}{\tau^2} \nn\\
 -\frac{(\delta.J)_{k'}}{\tau} & J_{kk'} & \frac{\phi_m (\delta.J)_{k'}}{\tau}
 \nn\\
 -\bphi_n \frac{C}{\tau^2} & \frac{\bphi_n (\delta.J)_{k'}}{\tau} &
 \bphi_n\phi_m
\frac{C}{\tau^2} + \frac{\threedx \delta_{nm}}{\tau} 
\end{array} \right) \, . 
\ee 
The inverse of this matrix is 
\be 
K^{\bbeta\alpha} = \left( 
\begin{array}{cc}
s^2 & 0 \nn\\
0 & K^{{\overline b} a} 
\end{array} \right) ,
\ee 
where 
\be
 K^{{\overline b} a}  = \frac{1}{\threedx}\left( 
\begin{array}{ccc}
\tau^2 + \tau \sum_n|\phi_n|^2  & \delta_k \tau & \phi_m \tau \nn\\
 \delta_{k'} \tau & A_{kk'} & 0 \nn\\
\bphi_n \tau   & 0 & \delta_{nm} \tau  
\end{array} \right) \, . 
\ee 
We have made no approximations to get this result. 
Notice that there is no mixing between the $M_k$ and the
$\phi_n$ fields.

The $F$-part of the scalar potential is given by 
\be
V_F=e^G (-3 + G_{\alpha} K^{\alpha \bbeta} G_{\bbeta}) = e^G B \, ,
\ee
where the reduced Planck mass is set to one, and where, 
assuming that the superpotential does not depend on $T$ 
(\ie the hidden sector group is on a D9-brane),
\ba
B &=& -3 + s^2 |G_S|^2 + K_T K^{ T \bT} K_{\bT} \nn\\
&&+ 2 Re 
\left( K_T K^{ T \bk} G_{\bk} +  K_T K^{ T \bn} G_{\bn} \right) \nn\\
&&+ G_{k'} K^{ k' \bk} G_{\bk} +  G_{n} K^{ n \bm} G_{\bm} \, .
\ea
Inserting the expressions for $K^{\alpha\bbeta}$ and 
completing the squares gives 
\ba
B & = & \delta.x - (\delta.A^{-1}.\delta) \threedx +  s^2 |G_S|^2 \nn\\
&& + \sum_n \left| \phi_n \sqrt{\frac{\threedx}{\tau}} - 
      G_{\bn}\sqrt{\frac{\tau}{\threedx}}\right|^2 \nn\\
&& + \sum_k \left| (\delta.A^{-\frac{1}{2}})_k \sqrt{\threedx} -
\frac{(G.A^{\frac{1}{2}})_k}{\sqrt{\threedx}}\right|^2  
\ea 
where $ (G.A^{\frac{1}{2}})_k=G_{k'}(A^{\frac{1}{2}})_{kk'}$ and
$(\delta.A^{-\frac{1}{2}})_k=\delta_{k'}(A^{-\frac{1}{2}})_{kk'}$.
Substituting for $K_T$, $G_n = K_n + W_n /W $, and $G_k = x_k + W_k / W$ we
finally get 
\be 
B = \delta.B_0 + s^2 |G_S|^2 + 
\frac{\tau}{\threedx}\sum_n \left| \frac{W_n}{W}\right|^2 
+ 
\sum_{k,k'}\left(B_{0\, k}+ \frac{W_k}{W}\right) 
\frac{A_{kk'} }{\threedx} 
\left(B_{0\, k'}+ \frac{\bW_{k'}}{\bW}\right) \, ,
\ee
where we have defined 
\be 
B_{0\, k} = x_k - (A^{-1}.\delta)_k \,\threedx \, .
\ee
This form of potential is obviously similar to the no-scale 
result, but there is a difference. Here part of the contribution 
to $B$ (\ie the $B_{0\, k}$ functions) can be negative and we
can (at least formally) have unbroken supersymmetry at 
finite values of parameters. To find the global minimum we 
set all the squares to zero which gives us 
\be 
B = \delta_k x_k 
- \frac{\delta_k \delta _{k'}}{(J^{-1})_{\,kk'} + 
\frac{\delta_k \delta_{k'}}{\threedx}} \, .
\ee
This function always has a minimum of $B=-3$ at $\delta.x =-3 $. 
Such large values of $x_k$ are almost certainly unphysical because 
$x_k$ (or $m_k$) describe the blowing up of the fixed points, and 
hence the orbifold `approximation' must break down.

We will mostly consider (for simplicity) only one $M_k$ field
which we call $M$ (it is easy to generalize to any number)
in which case 
\be  
B= \delta B_0 + \left|1-s\frac{W_S}{W}\right|^2 + 
\frac{\tau}{(3+\delta x)}\left| \frac{W_n}{W}\right|^2 
+  
\frac{A }{(3+\delta x)} 
\left| B_0+ \frac{W_M}{W}\right|^2 \, .
\ee
As well as this $F$-term contribution to the potential we have 
the $D$-term contribution 
\be
V_D=\frac{g_X^2}{2} |D_X|^2\, ,
\ee
where
\ba
D_X 
&=& \sum_n q_n K_n \phi_n + \delta _{GS_k} \frac{x_k}{2}
\nn\\
&=& \frac{\threedx}{\tau} q_n |\phi_n|^2 + \delta _{GS_k} \frac{x_k}{2}\, .
\ea
The Fayet--Iliopoulos term,
\be 
\xi^2=\frac{1}{2}\delta_{GS_k}x_k \, ,
\ee
is given by $m_k$ not $S$ and so can be zero. 

We now minimize the potential assuming that the final 
cosmological constant is small or zero  
(we will show this is possible later on)
so that we can ignore the first term in 
\[ V'_F = G' e^G B + e^G B' . \]
Assume that $\langle W \rangle \sim m_W$ (as 
phenomenology demands) so that we can impose the constraint 
$\langle D_X \rangle \approx 0$;
\be
q_2 |\phi_2|^2 + \sum_{n\neq 2}q_n |\phi_n|^2 =
\frac{|\delta_{GS}x|}{2}\frac{\tau}{(3+\delta x)}M_P^2\, .
\ee
For definiteness we will take $q_2 > 0$, $q_{n\neq 2}\leq 0$,
and here we are anticipating the fact that 
$\langle \delta_{GS} x \rangle$ will eventually be negative 
(see the end of this section).
On the other hand, the perturbative part of the superpotential
involves $\phi_{n\neq 2}$ and in order to make $\langle W\rangle \sim m_W$ 
we typically require
$|\phi_{n\neq 2}|^2\ll |\phi_2|^2$ and hence
\be
\label{fix2a}
q_2 \frac{|\phi_2|^2}{\tau}\approx\frac{|\delta_{GS}x|}{2(3+\delta x)}M_P^2\, .
\ee
As in the heterotic case, 
because of the smallness of $\phi_{n\neq 2}$ we can impose 
$W_{n\neq 2} = 0$, and since $W_{np}$ only involves $\phi_2$ 
gauge invariance of $W_p$ then tells us that 
\be 
\sum_n q_n W_{p\, n} \phi_n = 0 = W_{p\, 2}\, .
\ee

Let us briefly digress to discuss an explicit example of superpotential.
Consider a perturbative superpotential $W_p$ 
containing $\phi_2$ and two other fields $\phi_{1,3}$;
\be
W=a\phi_3(\phi_1\phi_2)+b(\phi_1\phi_2)^2 + \frac{c}{3}\phi_3^3 + W_{np}
\ee
Here we have taken $\phi_3$ to be a singlet, whilst $q_1=-q_2$. 
Then we have 
\bea
W_1&=&a \ \phi_2\phi_3 + 2b \ \phi_1\phi_2^2\\
W_2 &=&a \ \phi_1\phi_3 + 2b \ \phi_1^2\phi_2
+
\partial_2W_{np}=\frac{\phi_1}{\phi_2}W_1-
\frac{2}{N_c-1}\frac{W_{np}}{\phi_2}\\
W_3&=&a \ \phi_1\phi_2 + c\ \phi_3^2\, ,
\eea
and the solution to $W_1=W_{p\, 2}= W_3=0$ is 
\bea
\langle\phi_1\phi_2\rangle & = &-\frac{a^3}{4b^2c}\\
\langle\phi_3\rangle & = & \frac{a^2}{2bc}
\eea
so that
\be
\langle W_p \rangle=\frac{-a^6}{48b^3c^2}
\ee
The condition $m_{3/2}\sim m_{W}$ implies 
$\frac{a^6}{48b^3c^2}\sim 10^{-16}$. Since $b$ is dimensionful, 
it seems natural to associated the suppression with a large 
non-renormalizable term (coming from a low fundamental scale) 
giving a large $b$; \ie $b \sim 10^5 /M_P $. As promised, since $q_1$ 
is negative, imposing the $D_X=0$ condition then requires 
$\phi_2$ to be many orders of magnitude 
larger than $\phi_1$, for any reasonable value of $\tau$.

Returning now to our general discussion, we 
introduce the variable $\Delta=-\frac{W_S}{W}$ and rewrite $B$ in the
$\partial_n W_p=0$ directions:
\ba
\label{gnnn}
B 
&=&  \delta B_0 + \left|1+s\Delta\right|^2 +  
\frac{A }{(3+\delta x)} 
\left| B_0- \sigma \Delta\right|^2 + 
\frac{\tau}{(3+\delta x)}\left| \frac{\partial_2W_{np}}{W}\right|^2 
\nn\\
&=&  \delta B_0 + \left|1+s\Delta\right|^2 +  
\frac{A }{(3+\delta x)} 
\left| B_0- \sigma \Delta\right|^2 + 
\frac{q_2 }{8\pi^4|\delta_{GS}x|} \Delta^2
\, ,
\ea
where we have used
\be
\frac{\partial_2W_{np}}{W}=
-\frac{\Delta}{4\pi^2\phi_2}\, .
\ee
Note that $B_0$ only depends on $x$. 
Minimizing with respect to $s$ and $\Delta$ gives
\ba 
\label{solutions}
s & = & -\frac{1}{\Delta} \nn\\
\Delta & = & \frac{\sigma B_0}{\sigma^2 + \lambda^2} \,\, ; \, 0 \, ,
\ea
where we have defined 
\be
\lambda^2 = \frac{(3+\delta x)}{A}\frac{q_2}{8\pi^4|\delta_{GS}x|}\, .
\ee 
Note that the first solution always requires $B_0 \sigma < 0$.
The VEVs of $\phi_2$ and $\tau$ are determined by 
eqs.(\ref{fix2aa}) and (\ref{fix2a}) to be
\be
\phi_2^2 \approx \left(\frac{8\pi^2}{\Delta(N_c-1)
\bra W \ket}\right)^{N_c-1}e^{-8\pi^2S} \ \ \ \mbox{and} \ \ \
\tau \approx 
\frac{2 q_2 |\phi_2|^2 (3+\delta x)}{|\delta_{GS}x|} \, ,
\ee
so that, at this stage, $x$ is the only parameter remaining unfixed. 
Since $\langle W \rangle \sim e^{-4\pi^2 }$ in natural units, 
we can have virtually any value of $\phi_2$, with $\phi_2 \sim 1 $
corresponding to $N_c-1 \approx s$. 

The remaining potential is given by 
\ba 
B &=& B_0 \left(
\delta + \frac{A}{(3+\delta x)} \frac{\lambda^2 B_0}{\sigma^2+\lambda^2}
\right)\nn\\ 
&=& B_0 \frac{\delta }{\sigma^2+\lambda^2}
\left(
\sigma^2 - \frac{q_2}{8\pi^4 \delta_{GS}\delta}\right) \, .
\ea
There are now two options that one can consider in treating the 
remaining $x$ degree of freedom, and we now describe 
each of them in turn. 

\vspace{0.5cm}

\noindent{\underline{\it The `no-scale' case}}\\

\begin{center}
\begin{figure}[htp]
\begin{center}
\vspace*{-0.5in}
\epsfxsize=7.85in
\epsffile{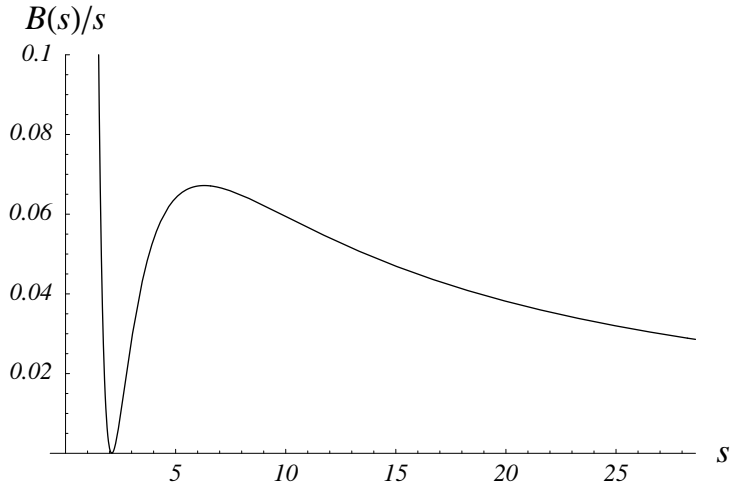}
\vspace{-6.5in}
\caption{Effective potential $B(s)/s$ with the 
cosmological constant set to zero; $s= {S+\bS}$} 
\label{v3}
\end{center}
\end{figure}
\end{center}

\noindent The first option is to set the cosmological constant to be exactly 
zero. As we have seen, $B_0$ must be non-zero to stabilize the 
dilaton, so instead we tune the parameter $\sigma$;
\be
\label{hmm}
\sigma^2 = \frac{q_2}{8 \pi^4 \delta\delta_{GS} }\, .
\ee
We should bear in mind that, since $\sigma $ depends on the particle
content, it is not a continuous parameter, and the
constraint can only be approximately satisfied by 
for example choosing $N_c$.
For the particular value of $\sigma$ in eq.(\ref{hmm}), the $m$ dependent 
VEVs we have found for $\phi_1$, $\phi_2$, $\tau$ and $s$ 
correspond to a flat direction in the parameter $m$ with 
zero cosmological constant. For generic small values of $x $ with 
$\delta_{GS}x < 0$ the dilaton is stabilized at
\ba
B_0(0) & \approx & - {\delta^2} \nn\\
s & \approx &  \frac{\sigma^2+\lambda^2}{\sigma \delta^2} \, .
\ea
As $m \rightarrow 0$, $\lambda$ dominates and 
the dilaton VEV diverges.
In figure (\ref{v3}), we plot the 
potential including the $1/s$ prefactor from $e^G$, 
with $\Delta=-1/s $ and imposing the cosmological constant condition. 
A natural possibility (which we will not explore here)
is that $x$ and therefore $m$ can be fixed (with $\delta_{GS}x < 0$), 
as in the conventional `no-scale' models, 
by minimizing the potential after radiatively induced 
electroweak symmetry breaking.

\noindent{\underline{\it The minimized $B_0(x)$ case}}\\


\noindent The second option is to tolerate a small but negative
cosmological constant which as we shall now see
allows a suitable local minimum in $x$. If we assume that 
$\lambda^2\ll \sigma^2 $ then we may simply minimize 
$\delta B_0$ which is given by 
\be 
\delta B_0 = \delta x - \frac{(3+\delta x)}{\frac{(3+\delta x)}
{\delta^2 J} + 1}\, .
\ee
In the case that $\sigma = \beta/2$
the assumption is true for large $N_c$.
$B_0$ is minimized where either $(3+\delta x) = 0 $ (\ie the 
unbroken supersymmetry minimum) or 
\be 
\label{mincond}
1+ \frac{ 2 \delta ^2 J}{(3+\delta x)} - \delta J_x  = 0 \, . 
\ee 
This can be satisfied for small $x$ by functions $J$ that vary 
sufficiently fast close to $x=0$. 
For small $x$ the extremum is always at 
\be 
\label{x}
x \approx \frac{1}{a \delta}\, ,
\ee
where $J = 1 + a \frac{x^2}{2} + \ldots \,$.
The sign of $B_{xx}$ is the same as that of $-J_{xx}$.
In other words for positive $a$ we get a maximum at small 
values of $x$ with $\delta_{GS}x >0$ and for negative $a$ we 
get a {\em minimum} at small values of $x$ with $\delta_{GS}x <0$.

\begin{center}
\begin{figure}[htp]
\vspace*{-0.7in}
\epsfxsize=7.85in
\epsffile{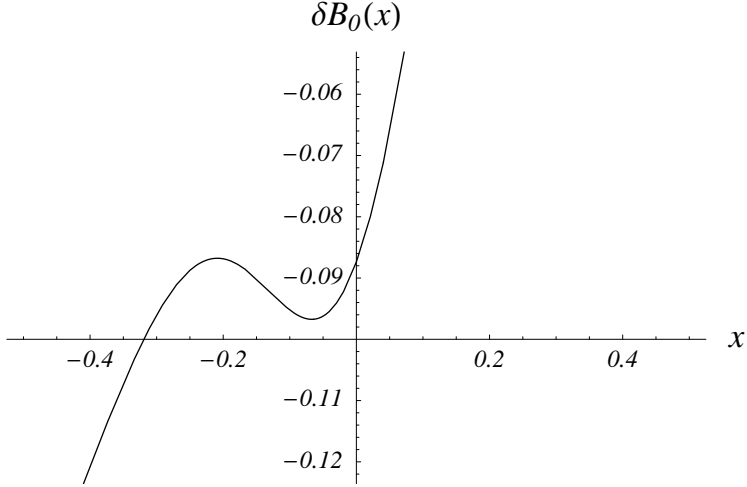}
\vspace{-6.5in}
\caption{The function $\delta B_0(x)$ for $J=e^{a x^2 /2}$ where $a=-60$.} 
\label{v2}
\end{figure}
\end{center}

We plot $B_0(x)$ in figure (\ref{v2}) for 
$J=e^{a x^2 /2}$ where 
$a=-60$ and $\delta=0.3$ (a large but 
realistic value according to ref.\cite{abd}). 
Any function $J$ that falls off sufficiently fast will form a minimum,
the important feature being 
the coefficient of $x^2 $ in $J$ or, since we are 
at small $x \sim m $, the value of $a = K''''$ at $m=0$ (where 
primes denotes differentiation with respect to $m$).
A perhaps more realistic example is 
\be 
\hat{K}(m) = \frac{3}{a} \ln\left( 1+ \frac{a}{6} m^2 \right)\, .
\ee
The minimum in $B_0$ is found at 
\be 
\delta B_0|_{min} \approx \frac{1}{2 a} - \delta^2 .
\ee
In other words, since $a$ is negative to form a minimum, 
the condition that $B_0\sigma < 0 $ to
stabilize the dilaton requires $\sigma \delta > 0 $.

All fields, including the $m$ field, can 
be stabilized with small negative 
cosmological constant provided we 
have a large negative value for $K''''(0)$.
Plugging our expressions for $\sigma$ and $B_0$
back into the solutions for $s$ in eq.(\ref{solutions}), we have
\be 
s \approx \frac{\sigma}{\delta^2}\, .
\ee

\bigskip 

To summarize, in both cases
the dilaton is stabilized primarily because of the 
$M/T$ mixing. Without this mixing (\ie setting $\delta=0$) we find 
only the runaway solution. The second important factor is
that the gauge kinetic functions
involve a linear combination of $S$ and $M$, described by the $\sigma$ 
coefficient which, like $\delta$, is model-dependent. 
We need a particular sign of 
$\sigma $ in order to have a minimum other than the usual runaway minimum. 

We also note that in the minimization there was an interesting interplay 
between the $D$-terms and the $F$-terms. 
This is similar to the $D$-term mediation of supersymmetry breaking
described in ref.\cite{bd}, except here different terms are 
set to zero at the minimum. (In the 
present case we have $W_1 \approx 0$ and $W_2\neq 0$ 
whereas in ref.\cite{bd} the minimization is at $W_2\approx 0$ 
and $W_1\neq 0$.) In order to achieve this we {\it a priori} need 
to choose a perturbative 
superpotential, $W_p$, which gets a non-zero expectation value.

\section{Supersymmetry breaking terms}

Since we have control over the VEVs of all the fields, 
it is now possible to write the complete expressions for
supersymmetry breaking without having to define an arbitrary
goldstino angle. Again we will consider only one $M_k$ field
for simplicity. The supersymmetry breaking effects are carried 
by the auxiliary fields $F^\alpha$
\be
F^\alpha =  e^{G/2 } G^{\alpha\bbeta} G_{\bbeta} \, .
\ee 
At the minimum,
\bea
\label{Wnp}
G_{\bS}&=&0 \nn\\
G_{\bM}&=&
x-B_0 \frac{\sigma^2}{\sigma^2+\lambda^2}
\nn\\
G_{\bn}&=&\frac{(3+\delta x)}{\tau} \phi_n \ \  \ \, \ n=1,3\nn\\
G_{\overline{2}}&=&\frac{(3+\delta x)}{\tau} \phi_2 
-\frac{\Delta}{4\pi^2\overline{\phi_2}} \, .
\eea
Defining 
\be 
\omega = \frac{\Delta}{4 \pi^2 (3+\delta x)}\, ,
\ee
this gives
\be
F^\alpha = \left( 0\, , \, F^T 
\, , \, F^M\, , \, F^i \right) 
\ee
where
\bea
F^T&=&e^{G/2} \tau \left(
\frac{\delta^2}{A}-1-
\omega \right)\nn\\
F^M&=& e^{G/2} \frac{\lambda^2}{\sigma^2+\lambda^2} \left(\frac{A}{(3+\delta
x)} x-\delta\right)\nn\\
F^1&=&F^3=0\nn \\
F^2&=& -e^{G/2}\frac{\tau}{\overline{\phi_2}}\omega \, .
\eea
The gaugino masses are given by 
\be 
M_a = \frac{1}{2} \left( Re (f_a) \right)^{-1} 
F^\alpha \partial_\alpha f_a 
\ee
where $f_a$ is the gauge kinetic function for the gauge group. 
For gauge groups that live on the D9-brane we have $f_a= S+\sigma_a M$ 
so that 
\be 
M_{9} = \frac{1}{2}m_{3/2} 
\frac{\sigma_a\lambda^2}{\sigma^2+\lambda^2}
\frac{\left(\frac{A}{(3+\delta x)}x-\delta\right)}{Re(S+\sigma_a M)} 
\ee
where the VEVs of $S$ and $M$ can be deduced from the 
expressions above. These relatively small D9 gaugino mass terms
arise solely due to the non-zero value $F^M$ in a manner 
suggested in ref.\cite{pheno1}. 

We shall present the remaining supersymmetry breaking 
in the limit where $\lambda^2\ll\sigma^2$, 
allowing us to drop terms of order $\lambda^2/\sigma^2 $
and to set $F^M=0$.
Consider the supersymmetry breaking for visible sector fields, 
$C_\alpha$, which occur in the same no-scale structure as 
the $\phi_1\, ,\phi_2$
fields (\ie they are $C^9_i$ or $C_i^{5j}$ fields). 
As usual, we expand the K\"ahler potential around $C_\alpha=0$
in a basis in which the K\"ahler metric is diagonal in the matter fields; 
\be
K = K_0 + 
{\cal Z}_{\alpha} |C_\alpha|^2 + \ldots
\ee
The expressions for the scalar masses (where $m_{3/2}=e^{G/2}$) are
\be
m_\alpha^2 = m_{3/2}^2 + V_0 - 
F^{\overline{\sigma}} F^\rho \partial_{\overline{\sigma}}
\partial_\rho \ln {\cal Z}_\alpha \, .
\ee
Substituting the K\"ahler potential we get 
\bea
|F^T|^2\partial_T \partial_{\bT}\ln {\cal Z}_\alpha 
&\approx & \frac{A^2(1+\omega -\frac{\delta^2}{A})^2}
{(A-\delta^2)^2} \, m_{3/2}^2 \nn\\
F^{\bT}F^2\partial_{\bT}\partial_2\ln {\cal Z}_\alpha 
& \approx & -\frac{A^2\omega}{(A-\delta^2)^2} 
(1+\omega-\frac{\delta^2}{A}) \, m^2_{3/2} \nn \\
|F^2|^2\partial_2 \partial_{\overline{2}}\ln {\cal Z}_\alpha
&\approx & \frac{A^2 \omega^2}{(A-\delta^2)^2} \, m^2_{3/2}\, ,
\eea
where we have used eq.(\ref{mincond}). Hence we find 
\be
m_{\alpha,9}^2 = V_0 \approx 0 \, .
\ee
This is a small negative mass-squared term of order $-\delta^2$.
However eq.(\ref{mincond}) is not true for the `no-scale' 
case, and also relies on our assumption that $\lambda^2\ll\sigma^2$.
If either of these conditions are not satisfied then we can get nett 
positive mass squared terms of order $\delta^2 m_{3/2}^2$.

Finally the $A$-terms for a Yukawa coupling $C_\alpha C_\beta C_\sigma$ 
are given by 
\ba
A_{\alpha\beta\gamma} 
& = & 
F^{\rho} 
\left( K_{\rho}+ \partial_{\rho}\ln Y_{\alpha \beta \gamma}  - \partial_{\rho} 
\ln
\left( 
{\cal Z}_\alpha {\cal Z}_\beta {\cal Z}_\gamma 
\right) \right) \nn\\
&=& 
m_{3/2}
\left(-\delta x+3\frac{\delta^2J}{3+\delta x}\right)\left(-1+\omega
\right) \nn\\ 
&\approx & 0 \, .
\ea
where we have assumed that 
$\partial_TY_{\alpha \beta \gamma}=\partial_2Y_{\alpha \beta \gamma}=0$.

So, for the 
fields and gauge groups associated with the D9-branes, the soft breaking
terms are suppressed by powers of $\delta^2 $. 
However, supersymmetry breaking can be 
more interesting for the fields living on the 
D5-branes. In general
the K\"ahler potential is of the form~\cite{imr}
\ba 
K &=& -\ln \left( s - \sum_i |C_i^{5_i}|^2  \right) - \sum_i 
\ln\left( \tau
 - |C_i^9|^2 
-  \sum_{j\neq k\neq i = 1}^3 |C_j^{5_k}|^2 \right) \nn\\
&& + \frac{1}{2}\sum_{j\neq k\neq i = 1}^3 
\frac{|C^{5_j\, 5_k}|^2}{s^{1/2} \tau^{1/2} }
+ \sum_{i = 1}^3 
\frac{|C^{95_i}|^2}{ \tau }
\, ,
\ea
where again we have assumed degenerate moduli fields $(T_i=T_j=T_k=T)$.
(As we mentioned in the introduction, 
our choice of putting the $\phi_n$ fields in a single $\tau$ 
formally corresponds to the linear combination $\phi_n=
\frac{1}{\sqrt{3}}(C^9_{1\, n}+C^9_{2\, n}+C^9_{3\,n })$. 
However, once we have imposed $D_X=0$, 
the VEVs of $\phi_n$ and $\tau$ indicate that
the K\"ahler potential we have been using 
throughout is equivalent to the above 
upto order $(\delta_{GS} x)^2$, independently of the particular 
linear combination.). This gives us the following supersymmetry 
breaking pattern. \\

\vspace{0.2cm}

\noindent {\underline{\it Mass-squareds}};
\ba
C_i^9 \, , \, C^{5_i}_{j\neq i} \, , \, C^{95_i} 
 \, &:& \, m^2_\alpha = m^2_{\alpha,9} \approx 0\nn\\
C^{5_i5_j} \, &:& \, 
m^2_\alpha 
\approx \frac{1}{2} m_{3/2}^2 \nn\\
C_i^{5_i} \, &:& \, m^2_\alpha \approx m_{3/2}^2 \, .
\ea


\noindent {\underline{\it Gaugino Masses}};
\ba
M_9 &\approx & 0 \nn\\
M_{5_i \, a} 
&\approx & -\frac{1}{2}m_{3/2} \frac{\tau}{Re(T+\sigma_a M))} \, . 
\ea


\noindent {\underline{\it $A$-terms}};

\vspace{0.2cm}

\be
A_{\alpha\beta\gamma} 
\approx m_{3/2} \left( 3 -(\alpha +\beta +\gamma)\right)
\ee
\vspace{0.2cm}
where 
\be 
\alpha =-\tau \partial_T\ln{\cal Z}_{\alpha}= \left( 1 \, , \, \frac{1}{2}\, ,0 \, \right)
\,\,\, \mbox{ for } \,\,\, \left(
C_i^9 \, C^{95_i} \, C^{5_i}_{j\neq i} \,\, , \,\, 
 C^{5_i5_j} \,\, , \,\, C_i^{5_i} \ \right)\,\, \ \mbox{respectively},
\ee
so that, for example, a coupling between $C_i^9 C^{5_i5_j} C_i^{5_i} $ 
would give $A=3/2 m_{3/2}$. 
Note the usual sum-rule relating $A$-terms and 
mass-squareds~\cite{imr}; 
\be 
m_\alpha^2 +m_\beta^2 + m_\sigma^2 = A_{\alpha\beta\gamma} \, .
\ee

To conclude, while the auxiliary fields $F^2$ can be larger than 
$F^T$, under the simplifying 
assumptions we have made, the soft masses and $A$-terms
are independent of $F_2$. In addition 
the supersymmetry breaking shows a rather interesting structure which 
may allow us to realise of a number of suggestions that have been
put forward as solutions to the supersymmetric flavour and
CP problems. For example it might be
possible to make the first and second generations of squarks 
heavy \cite{giudice} 
(of order a few TeV) if they are $C^{5i}_{i}$ fields 
whilst the 3rd generation is a $C^{5i}_{j\neq i}$ field. Alternatively, 
if the higgs plus first generation particles are
$C^{5i}_{i\neq j}$ fields, one would have an 
interesting non-universal structure for the 
$A$-terms~\cite{khalil}, 
and a suppression of contributions to electric dipole moments, 
in a manner similar to that described in ref.\cite{susycp}.

\section{Conclusion}

In this paper we considered gaugino condensation in 4D effective
theories of type I strings, and 
discussed its effect on dilaton stabilization and the structure of
soft breaking terms. Our main observation was that dilaton 
stabilization is possible with only one gaugino condensate.
An important role is played by the twisted ($M$) moduli fields 
which are a novel feature of these models. 
These fields enter in two important ways;
\begin{itemize}
\item
First they contribute a new term to the K\"ahler potential,
which includes, at the one-loop level, a mixing with the $T$-moduli. 
This mixing is one essential feature 
preventing the dilaton running away to infinity as in the heterotic case.
\item The other crucial ingredient comes from the tree level
$M$-dependance in the gauge kinetic functions. This
leads to some unusual properties of the non-perturbative dynamics of
these gauge theories. 
In particular the $M$ field appears in  
the condensation scale and it is this, 
combined with the $M/T$ mixing 
in the K\"ahler metric, that can stabilize the 
dilaton.
\end{itemize}
We found that dilaton stabilization occurs 
quite generally with two possible outcomes. 
In the first we set the cosmological constant to be exactly zero and 
all fields except the $M$ field are fixed. The $M$ field 
then parameterizes the supersymmetry 
breaking in a way which is reminiscent of `no-scale' models. 
The second possibility is to tolerate a small negative cosmological
constant. In this case we showed that certain types of
$M$ dependent terms in the K\"ahler potential can lead to a
stabilization of all fields including the $M$ field itself.
(The latter is stabilized at values close to the orbifold point.)
Although there is still some ignorance about the precise form of the 
$M$ dependence in the K\"ahler potential,
we were able to derive the general  
conditions required to develop a minimum for $M$;
namely that $\partial^4 K/\partial M^4$ must be large and negative.

The issues we have presented here certainly deserve further
investigation. For example, the phenomenological possibilities of 
the resulting supersymmetry breaking patterns seem promising
and we briefly mentioned some potential avenues of exploration.
Furthermore, in this paper we have made only the simplest 
(in a sense, minimal) assumption, that the condensing 
gauge group lives on a D9-brane. 
It would be interesting to consider cases in which 
the gauge groups and mesons are assigned differently.
In addition we have not touched upon the question of 
fundamental scales; it may be interesting to re-examine ideas 
such as `mirage
unification', in which the apparent unification  
of couplings is partly explained by the VEV of $M$~\cite{mirage}. 
Consistent mirage unification would directly relate 
the unification scale to the parameters in the theory (\ie $\delta$ and 
$\sigma$) that determine $\langle M \rangle$.

\section*{Acknowledgements}

We would like to thank Emilian Dudas, 
Shaaban Khalil, St\'ephane Lavignac, Fernando Quevedo, Carlos Savoy 
for enlightening discussions and suggestions, and for generous support. 


\end{document}